\documentclass[sigconf,9pt,screen]{acmart}
\renewcommand\footnotetextcopyrightpermission[1]{}
\usepackage{multirow}
\usepackage{tikz}
\usetikzlibrary{positioning, fit, backgrounds, arrows.meta, calc}
\usepackage{pgfplots}
\pgfplotsset{compat=1.18}
\usepackage{pgfplotstable}
\usepgfplotslibrary{groupplots}
\usepgfplotslibrary{statistics}
\usepackage{makecell}
\usepackage[font=footnotesize,skip=1pt]{caption}
\usepackage{subcaption}
\usepackage{algorithm}
\usepackage{algpseudocode}
\usepackage{braket}

\usepackage[table]{xcolor}

\definecolor{lossred}{RGB}{255,235,238}
\definecolor{nearred}{RGB}{255,205,210}
\definecolor{neutralyellow}{RGB}{255,249,196}
\definecolor{winlight}{RGB}{232,245,233}
\definecolor{winmid}{RGB}{165,214,167}
\definecolor{windark}{RGB}{76,175,80}

\AtBeginDocument{%
  }

\setcopyright{none}
\copyrightyear{2026}
\acmYear{2026}
\acmConference[ICCAD '26]{IEEE/ACM International Conference on Computer-Aided Design}{November 8--12, 2026}{San Jose, CA, USA}

\begin{document}

\title{\emph{VQCSim}: When Does Compile-Once Statevector Simulation Beat Generic Quantum Frameworks?}

\author{Anton Firc\textsuperscript{1}, Martin Peresini\textsuperscript{1}, Vojtech Mrazek\textsuperscript{2}, Kamil Malinka\textsuperscript{1}, Vojtech Stanek\textsuperscript{1}, Zbynek Licka\textsuperscript{1}, Nouhaila Innan\textsuperscript{3,4}, Walid El Maouaki\textsuperscript{3}, Alberto Marchisio\textsuperscript{3,4}, Muhammad Shafique\textsuperscript{3,4}}
\affiliation{%
  \institution{\textsuperscript{1}Security@FIT, Brno University of Technology, Brno, Czech Republic\\
  \textsuperscript{2}EvoAI Hardware Group, Brno University of Technology, Brno, Czech Republic\\
  \textsuperscript{3}eBRAIN Lab, Division of Engineering, New York University Abu Dhabi (NYUAD), Abu Dhabi, UAE\\
\textsuperscript{4}Center for Quantum and Topological Systems (CQTS), NYUAD Research Institute, NYUAD, Abu Dhabi, UAE\\
  }
  \country{}}
\email{{ifirc,iperesini,mrazek,malinka,istanek,ilicka}@fit.vut.cz}
\email{{nouhaila.innan,walid.el.maouaki,alberto.marchisio,muhammad.shafique}@nyu.edu}

\renewcommand{\shortauthors}{Firc et al.}

\begin{abstract}
Hybrid quantum--classical machine learning workflows repeatedly evaluate many small parametrized circuits during training and model exploration. In this regime, framework dispatch and orchestration overhead often dominate runtime. Prior simulators accelerate execution but leave open the question of \textit{when} compile-once specialization is the right choice for static variational circuits.
We answer this question with \textit{VQCSim}, a compile-once, PyTorch-native statevector execution path with native \texttt{autograd}. In a systematic MQT~Bench study, VQCSim compiles all tested static circuits and provides \(87.7\%\) end-to-end semantic validation. Across a five-GPU evaluation set, VQCSim delivers pooled median speedups of \(4.49\times\) for native inference and \(26.78\times\) for native training, while retaining a \(3.31\times\) advantage under matched finite-difference training. Ablation identifies native autograd as the dominant source of acceleration (\(27.6\times\)), with compile-once caching and batch vectorization contributing additional gains.
The speedup trades higher GPU memory (VQCSim is memory-limited at the high end) for lower runtime. We derive a hardware-aware regime map and release \texttt{vqcsim-oracle}, an open-source backend selector with \(91.1\%\)--\(97.7\%\) top-1 agreement (including cross-GPU transfers), enabling automatic simulator selection in QML design loops.
\end{abstract}



\maketitle

\section{Introduction}
\label{sec:intro}

Hybrid quantum machine learning (QML) workflows repeatedly evaluate small parametrized quantum circuits during training, ablation, and model selection loops. Because present-day quantum hardware remains limited in qubit count, noise, and runtime, much of this development is still performed in classical simulation, with ideal statevector simulation serving as the dominant exact reference for small systems~\cite{PERALGARCIA2024100619,gujju_survey_2024,zaman2025surveyquantummachinelearning,Qandle,TQml}. Recent QML studies routinely operate in the few-to-low-teens qubit regime after feature reduction, compact encoding, or hybrid preprocessing, and report useful results with circuits in the 4--16 qubit range~\cite{Jones2026,kolle2025,franco2025quantumphasesclassificationusing,Ara2025}.

In this regime, end-to-end cost is often driven not only by statevector arithmetic, but also by repeated framework-side circuit construction, dispatch, and orchestration overhead~\cite{Qandle,TQml}. Prior QML-oriented simulators demonstrate acceleration in this setting, but they do not resolve the central computer-aided design (CAD) question of \textit{when compile-once specialization is the right choice for static quantum circuits: which circuits admit it, which ones benefit most, and how the tradeoff between runtime and GPU memory shifts relative to a general-purpose simulator}~\cite{Qandle,TQml,quetschlich2023mqtbench}.

\begin{figure}[t!]
\centering
\includegraphics[width=\linewidth]{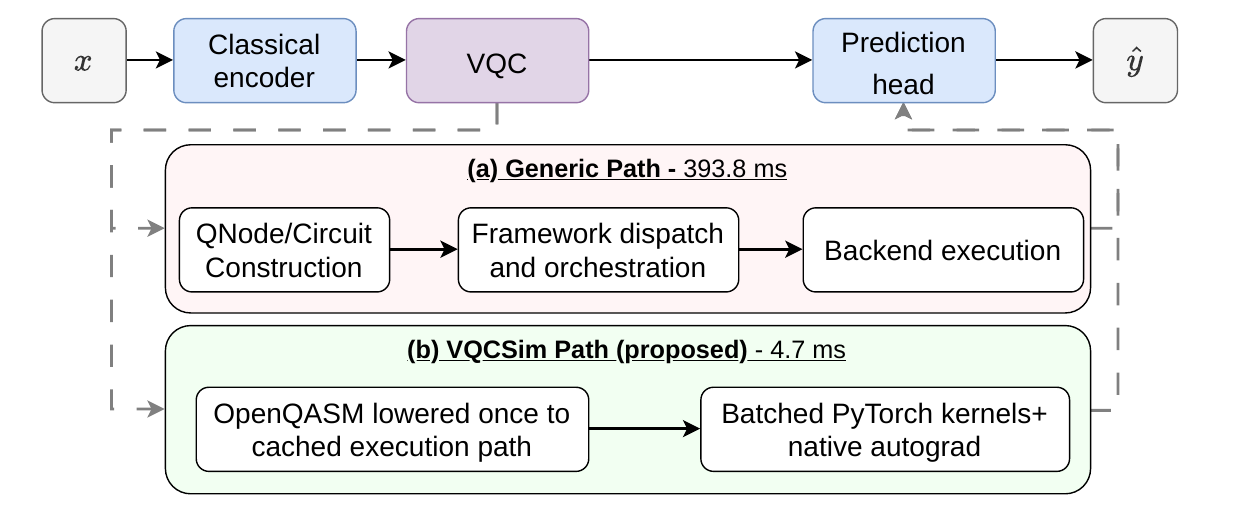}
\caption{Execution-path comparison for the VQC block in a hybrid quantum-classical model. The generic path performs circuit construction, framework dispatch, and backend execution, whereas VQCSim uses a cached OpenQASM-derived PyTorch execution path with batched kernels and \texttt{autograd} differentiation. The representative inference case shown (QNN, size~10, batch~128) illustrates the overhead eliminated by VQCSim.}
\label{fig:graphical_abstract}
\vspace*{-2em}
\end{figure}

We address this question with \textit{VQCSim}, a compile-once Torch-native execution path for supported static variational quantum circuits under ideal statevector simulation. \textit{VQCSim} is scoped to the small-to-moderate qubit regimes typical of current QML workloads and is positioned primarily as a practical CAD decision aid rather than solely as a new simulator. \autoref{fig:graphical_abstract} illustrates the central idea: the surrounding hybrid model remains unchanged, while the variational quantum circuit (VQC) block is redirected from a generic framework execution stack to a cached Torch-native path implemented in PyTorch.

Our main contributions are:
\begin{itemize}
    \item A compile-once Torch-native lowering path for static VQC-style circuits, validated at \(87.7\%\) end-to-end support on static MQT~Bench circuits~\cite{quetschlich2023mqtbench}.
    \item A systematic performance characterization revealing batch-invariant scaling in VQCSim versus batch-amplifying scaling in generic frameworks. Pooled median speedups reach 4.49$\times$ (inference) and 26.78$\times$ (training), with a 3.31$\times$ advantage under matched finite-difference training.
    \item An internal ablation identifying autograd as the dominant speedup mechanism (\(27.6\times\)), with compile-once caching and batch vectorization as complementary contributors.
    \item A hardware-aware regime map and open-source backend selector (\texttt{vqcsim-oracle}) achieving 91.1\%--97.7\% top-1 agreement in cross-GPU transfer.
\end{itemize}

Although prior simulators such as TorchQuantum~\cite{wang2022torchquantum}, Qandle~\cite{Qandle}, and TQml~\cite{TQml} explore pieces of this design, including compile-once lowering, Torch-native execution, and native autograd, they do not provide the full CAD-oriented picture. VQCSim complements these ideas with a systematic applicability study of which circuit classes can be lowered, an analysis of the batch-invariant and batch-amplifying regimes that determine when specialization is beneficial, and a validated backend selector that automates this choice. This makes VQCSim a practical CAD component for hybrid QML workflows, not just a faster simulator.

All code, including VQCSim and the open-source decision tool \texttt{vqcsim-oracle}, is publicly released for reproducibility and adoption at \url{https://github.com/Security-FIT/VQCSim}.

\section{Related Work}
\label{sec:related-work}

\noindent\textbf{QML workloads and exact simulation.}
Current QML development remains heavily constrained by limited and noisy quantum hardware, keeping much of model development, debugging, and benchmarking in classical simulation~\cite{PERALGARCIA2024100619,gujju_survey_2024,zaman2025surveyquantummachinelearning}. Ideal statevector simulation serves as the standard exact reference for small systems and is routinely used in hybrid QML studies operating in the low-to-moderate qubit regime~\cite{Jones2026,kolle2025,franco2025quantumphasesclassificationusing,Ara2025}.

\noindent\textbf{Quantum circuit simulators and GPU acceleration.}
General-purpose statevector simulators span a wide performance range~\cite{kharsa2024benchmarking}. Qulacs provides fast C++ simulation with optional GPU acceleration~\cite{suzuki2021qulacs}. NVIDIA's cuQuantum SDK provides CUDA-accelerated state-vector and tensor-network backends~\cite{cuquantum2023}. BQSim~\cite{jiang2025bqsim} accelerates batch circuit simulation on GPU via decision-diagram-based gate fusion, achieving 3--300$\times$ speedups over cuQuantum on deep circuits, but targets general simulation rather than QML-specific workloads. CAST provides a cross-platform compilation toolchain with sparsity-aware gate fusion for Schr\"odinger-style si\-mu\-la\-ti\-on~\cite{lu2025versatilecrossplatformcompilationtoolchain}. A recent concurrent framework performs empirical runtime backend selection across CuPy, PyTorch-CUDA, and NumPy backends~\cite{kumaresan2026gpuaccel}.

\noindent\textbf{QML-oriented simulator optimization.}
TorchQuantum~\cite{wang2022torchquantum} provides a PyTorch-native framework with GPU-accelerated statevector simulation, native \texttt{autograd}, and batched execution, and has been applied to quantum circuit robustness estimation and quantum neural architecture search. Qandle accelerates PyTorch-compatible statevector simulation via gate-matrix caching and circuit splitting~\cite{Qandle}. TQml optimizes simulation at the layer level by selecting per-layer primitives as a function of gate structure and qubit count~\cite{TQml}. Qiskit-Torch-Module provides a PyTorch wrapper for Qiskit-based quantum neural networks with reduced prototyping overhead~\cite{meyer2024qiskittorch}.

\noindent\textbf{Search-time evaluation and quantum architecture search.}
Quantum architecture search (QAS) requires repeated evaluation of many candidate ans\"atze, making fast inner-loop simulation valuable~\cite{du2022qas,martyniuk2024qas,dutta2025qas}. This pattern mirrors established CAD search workflows that use efficient evaluators in the inner loop and higher-fidelity tools for validation~\cite{mrazekSim2017,rohnas2022,nascaps2020}.

VQCSim differs from these systems in scope and methodology. Although prior work has explored building blocks such as compile-once lowering, Torch-native execution, and native autograd, VQCSim adds what those systems do not: a systematic MQT~Bench applicability study of which circuit classes can be lowered, a characterization of the scaling regimes that determine when specialization pays off, mechanism-level ablation, and a validated hardware-aware backend selector. This makes VQCSim a practical CAD component for hybrid QML workflows, not just a faster simulator. We compare directly against PennyLane, Qiskit, TensorCircuit, and CUDA-Q as representative general-purpose frameworks. TQml was excluded because its current offering is proprietary and access-controlled rather than publicly reproducible in our evaluation setting.

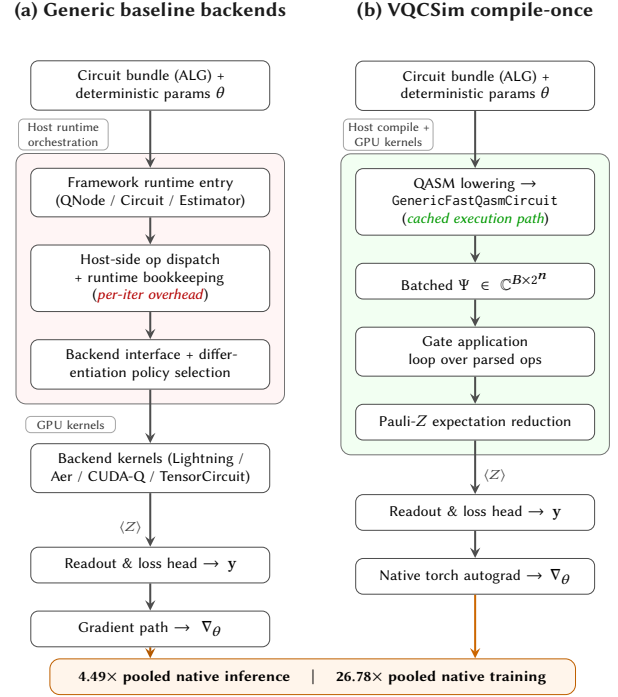
\begin{figure}[t]
\centering
\resizebox{0.95\linewidth}{!}{%
\begin{tikzpicture}[
    >=stealth,
    node distance=0.38cm and 0.75cm,
    basic/.style={draw=black!70, fill=white, rectangle, rounded corners=3pt,
                  minimum height=0.50cm, text width=3.1cm, align=center,
                  font=\sffamily\scriptsize, inner sep=3.5pt},
    title/.style={font=\sffamily\small\bfseries, text=black!90},
    box_slow/.style={draw=black!55, fill=red!4, rectangle, rounded corners=4pt, inner sep=0.20cm},
    box_fast/.style={draw=black!55, fill=green!5, rectangle, rounded corners=4pt, inner sep=0.20cm},
    lbl/.style={font=\sffamily\tiny, text=black!75},
    banner/.style={draw=orange!75!black, fill=orange!8, rectangle, rounded corners=4pt,
                   minimum height=0.50cm, text width=7.15cm, align=center,
                   font=\sffamily\scriptsize\bfseries},
    chip/.style={draw=black!35, fill=white, rectangle, rounded corners=2pt,
                 inner sep=1.5pt, font=\sffamily\tiny, text=black!80,
                 align=center, text width=1.15cm},
    arrow/.style={->, draw=black!70, thick}
]

\node[title] (titleA) at (0,0) {(a) Generic baseline backends};
\node[title] (titleB) at (4.55,0) {(b) VQCSim compile-once};

\node[basic, below=0.45cm of titleA] (a_fe) {Circuit bundle (ALG) + deterministic params $\theta$};

\node[basic, below=0.80cm of a_fe] (a_qn) {Framework runtime entry\\(QNode / Circuit / Estimator)};
\node[basic, below=of a_qn] (a_disp) {Host-side op dispatch + runtime bookkeeping\\(\textcolor{red!70!black}{\textit{per-iter overhead}})};
\node[basic, below=of a_disp] (a_orch) {Backend interface + differentiation policy selection};
\node[basic, below=0.75cm of a_orch] (a_back) {Backend kernels (Lightning / Aer / CUDA-Q / TensorCircuit)};
\node[basic, below=0.75cm of a_back] (a_read) {Readout \& loss head $\to \mathbf{y}$};
\node[basic, below=of a_read] (a_grad) {Gradient path $\to \nabla_\theta$};

\node[basic, below=0.45cm of titleB] (b_fe) {Circuit bundle (ALG) + deterministic params $\theta$};

\node[basic, below=0.80cm of b_fe] (b_sv) {QASM lowering $\to$ \texttt{GenericFastQasmCircuit}\\(\textcolor{green!60!black}{\textit{cached execution path}})};
\node[basic, below=of b_sv] (b_rot) {Batched $\Psi \in \mathbb{C}^{B\times 2^n}$};
\node[basic, below=of b_rot] (b_ring) {Gate application loop over parsed ops};
\node[basic, below=of b_ring] (b_pauli) {Pauli-$Z$ expectation reduction};
\node[basic, below=0.75cm of b_pauli] (b_read) {Readout \& loss head $\to \mathbf{y}$};
\node[basic, below=of b_read] (b_grad) {Native torch autograd $\to \nabla_\theta$};

\begin{scope}[on background layer]
    \node[box_slow, fit=(a_qn) (a_disp) (a_orch)] (slowbox) {};
    \node[box_fast, fit=(b_sv) (b_pauli)] (fastbox) {};
\end{scope}

\draw[arrow] (a_fe) -- (a_qn);
\draw[arrow] (a_qn) -- (a_disp);
\draw[arrow] (a_disp) -- (a_orch);
\draw[arrow] (a_orch) -- (a_back);
\draw[arrow] (a_back) -- node[lbl,left,pos=0.65] {$\langle Z \rangle$} (a_read);
\draw[arrow] (a_read) -- (a_grad);

\draw[arrow] (b_fe) -- (b_sv);
\draw[arrow] (b_sv) -- (b_rot);
\draw[arrow] (b_rot) -- (b_ring);
\draw[arrow] (b_ring) -- (b_pauli);
\draw[arrow] (b_pauli) -- node[lbl,right,pos=0.65] {$\langle Z \rangle$} (b_read);
\draw[arrow] (b_read) -- (b_grad);

\node[banner, below=0.78cm of $(a_grad)!0.5!(b_grad)$] (speed)
{4.49$\times$ pooled native inference \quad $|\quad$ 26.78$\times$ pooled native training};

\draw[->, thick, dashed, orange!70!black] (a_grad) -- (a_grad |- speed.north);
\draw[->, thick, orange!80!black] (b_grad) -- (b_grad |- speed.north);

\node[chip, anchor=south west] at ([xshift=1pt,yshift=1pt]slowbox.north west) {Host runtime\\orchestration};
\node[chip, anchor=south west] at ([xshift=-2pt,yshift=4pt]a_back.north west) {GPU kernels};
\node[chip, anchor=south west] at ([xshift=1pt,yshift=1pt]fastbox.north west) {Host compile +\\GPU kernels};

\end{tikzpicture}
}
\vspace*{1em}
\caption{Generic baselines incur framework/runtime orchestration and repeated host dispatch before backend kernels execute. VQCSim compiles OpenQASM once into a cached Torch-native execution path, and runs batched statevector kernels with native \texttt{autograd}.}
\label{fig:architecture}
\vspace*{-2em}
\end{figure}

\section{VQCSim Design}
\label{sec:method}

\textit{VQCSim} is a generic Torch-native quantum execution backend for supported static circuits. As shown in \autoref{fig:architecture}, it compiles OpenQASM once into a cached execution path and then executes the resulting circuit through differentiable PyTorch-based statevector simulation.

\noindent\textbf{Static Circuit Scope.}
VQCSim targets \emph{static} quantum circuits: circuits whose gate sequence and wire connectivity are determined at compile time. A circuit is static if it has no mid-circuit measurement, no classically controlled gates, and no data-dependent gate selection. Parametrized rotation gates (e.g., $R_z(\theta)$, $R_y(\theta)$) are permitted: only the \emph{angle} varies at runtime; gate identity and wiring stay fixed. This scope covers the variational ans\"atze common in hybrid QML workflows (QAOA, VQE, QNN layers) but excludes dynamic control flow such as repeat-until-success or adaptive measurement.

\noindent\textbf{Compile-Once Lowering and Execution.}
The key idea is compile-once, execute-many execution. A normalized OpenQASM circuit is lowered once into an internal gate sequence, where gates with constant numeric parameters are precomputed and cached as static unitary matrices, and parametrized gates remain symbolic. At runtime, VQCSim resolves the current parameter values, initializes a batched state vector $\Psi \in \mathbb{C}^{B \times 2^n}$ from either $\ket{0}^{\otimes n}$ or a provided initial state, and applies the gate sequence directly in Torch. This avoids repeated framework-level circuit dispatch and keeps the full quantum path inside standard Torch kernels.



\noindent\textbf{Gate Application and Differentiation.}
Gate application is implemented by reshaping the batched state, permuting the acted-on wires to contiguous trailing axes, applying the local operator with batched \texttt{torch.matmul}, and restoring the original wire order (\autoref{alg:vqcsim}). From the final state, VQCSim returns basis-state probabilities or Pauli-$Z$ expectations, while gradients with respect to trainable circuit parameters are obtained through native \texttt{autograd}.

\noindent\textbf{Complexity and Tradeoffs.}
The compile phase runs once in $\mathcal{O}(G)$ time, where $G$ is the gate count. Each subsequent forward pass costs $\mathcal{O}(B \cdot G \cdot 2^n)$ time with $\mathcal{O}(B \cdot 2^n)$ memory for the state vector, where $B$ is the batch size and $n$ is the qubit count. During training, PyTorch's autograd retains intermediate states, increasing memory to $\mathcal{O}(B \cdot G \cdot 2^n)$. The exponential scaling in $n$ is the fundamental constraint that defines VQCSim's applicability boundary. VQCSim therefore targets relatively small but heavily reused circuits, especially batched hybrid workloads where reducing framework overhead is more important than extending the absolute qubit limit.

\section{Research Questions and Experimental Design}

We evaluate VQCSim through four research questions that address applicability, performance regimes, bottlenecks, and practical simulator selection~\cite{quetschlich2023mqtbench,Qandle,TQml}.

\subsection{Research Questions}
\label{subsec:research_questions}


\noindent \textbf{\textit{RQ1 (Applicability):}} How broadly can a compile-once Torch-native execution path be constructed for MQT Bench circuits without changing circuit semantics?

\noindent \textbf{\textit{RQ2 (Performance):}} For which static quantum circuits does the specialized Torch-native path outperform generic simulators, and how do execution time and peak GPU VRAM scale with circuit descriptors?

\noindent \textbf{\textit{RQ3 (Mechanisms):}} Which components dominate runtime and memory in VQCSim and in generic simulators, and what mechanisms create scaling limits?

\noindent \textbf{\textit{RQ4 (Selection):}} Given circuit features and hardware constraints, when should a user prefer the VQCSim over a generic simulator?


\begin{algorithm}[t]
\caption{VQCSim: Compile-Once Lowering and Batched Execution}
\label{alg:vqcsim}
\begin{algorithmic}[1]
\Require normalized OpenQASM circuit $C$, parameter vector $\theta$, batch size $B$, optional initial state $\Psi_0$
\Ensure Pauli-$Z$ expectation matrix $E \in \mathbb{R}^{B \times n}$

\Statex \textbf{--- Compile phase (executed once, result cached) ---}
\State Parse $C$ into gate sequence $\mathcal{G}=\{(g_i,w_i,p_i)\}_{i=1}^{G}$
\State For each \textit{constant} gate $g_i$: precompute and cache matrix $U_i \in \mathbb{C}^{2^{|w_i|}\times 2^{|w_i|}}$
\State Classify each gate as \textsc{static} (cached) or \textsc{parametric} (runtime)

\Statex \textbf{--- Execute phase (per call, parameters $\theta$ may vary) ---}
\State Initialize $\Psi \gets \Psi_0$ if provided, else $\Psi \gets |0\rangle^{\otimes n}$ tiled to shape $[B,\,2^n]$
\State \textbf{for} $(g_i,w_i,p_i)$ in $\mathcal{G}$ \textbf{do}
\State \hspace{1.6em}\textbf{if} $g_i$ is \textsc{static} \textbf{then} retrieve cached $U_i$
\State \hspace{1.6em}\textbf{else} instantiate $U_i$ from current $\theta$
\State \hspace{1.6em}\textbf{end if}
\State \hspace{1.6em}Apply $U_i$ to $\Psi$ on wires $w_i$ via reshape/permutation and batched \texttt{torch.matmul}
\State \textbf{end for}
\State \textbf{return} $E_{:,j}=\sum_k (-1)^{k_j}\,|\Psi_{:,k}|^2$ for each qubit $j$
\end{algorithmic}
\end{algorithm}

\subsection{Experimental Setup}
\label{subsec:experimental_setup}

Experiments run on Linux with five NVIDIA GPUs (RTX~5080, RTX~4000 Ada, RTX~4500 Ada, RTX~A5000, RTX~A6000). The main evaluation and ablation runs use the RTX~5080; cross-GPU results in RQ2 are reported on the common five-GPU subset. The software stack is based on PyTorch~2.10.0 with Qiskit, PennyLane Lightning, TensorCircuit, and CUDA-Q~\cite{asadi2024lightning,tensorcircuit2023,qiskit2024,cudaquantum}. Per-run manifests record the full environment (Python, Torch, CUDA, cuDNN, and framework versions, GPU model, and total VRAM).

\subsubsection{Benchmarks and Baselines}
\label{subsubsec:benchmarks_baselines} 

All circuits are generated from \texttt{mqt.bench} through the Python API at \texttt{BenchmarkLevel.ALG} from a pinned installed release \cite{quetschlich2023mqtbench}. Mirror-circuit generation and random-parameter generation are disabled throughout. Benchmark families are enumerated from the pinned release, and circuit sizes are discovered separately for each family up to a size of 20. Dynamic families are tracked explicitly but excluded from static-support accounting.

RQ1 uses the full set of discovered static MQT Bench instances up to the configured size limit. Its baseline is a semantic reference rather than a competing performance backend, so correctness is evaluated against Qiskit execution after lowering~\cite{qiskit2024}.

RQ2 uses the variational families \texttt{qaoa}, \texttt{qnn}, \texttt{vqe\_real\_amp}, \texttt{vqe\_ su2}, and \texttt{vqe\_two\_local}, generated under the same fixed policy~\cite{quetschlich2023mqtbench}. We compare VQCSim against PennyLane, Qiskit, CUDA-Q, and TensorCircuit \cite{bergholm2018pennylane,qiskit2024,tensorcircuit2023,cudaquantum}. Each circuit is evaluated in inference and training modes with batch sizes of 1, 8, 32, and 128. CUDA-Q is included only for inference because the evaluated path does not provide the same training interface used for the other frameworks.

\subsubsection{Preprocessing and Semantic Validation}
\label{subsubsec:preprocessing_validation}

When a circuit is parametric, its parameters are ordered deterministically and bound to a fixed parameter vector. For RQ1, each bound circuit is then classified by dynamic versus static behavior, measurement pattern, gate support before and after preprocessing, and a coarse structural class. For both RQ1 and RQ2, circuits are decomposed and transpiled with a fixed Qiskit policy: a fixed supported-basis gate set, \texttt{optimization\_level=0}, \texttt{coupling\_map=None}, and a fixed \texttt{seed\_ transpiler}~\cite{qiskit2024}.

For RQ1, we test whether each preprocessed circuit can be lowered into the generic Torch-native execution path. When lowering succeeds, we validate the result against a Qiskit reference using measurement-aware parity checks~\cite{qiskit2024}. For circuits without terminal measurement, validation uses state fidelity, maximum statevector difference up to global phase, maximum probability difference, and an additional \texttt{Statevector.equiv(...)} sanity check. For circuits with terminal computational-basis measurement, validation is based on probability parity. For each circuit, we record whether lowering succeeded, whether validation passed, and, if lowering failed, the first operation or construct that prevented support. We then aggregate these outcomes into overall coverage statistics.

For performance analysis (RQ2), each framework-specific prepared circuit is parity-checked against a Qiskit reference before timing. When a backend exposes a state vector, parity uses the same state-based checks; when only expectation values are available, parity falls back to expectation-value agreement against the same reference. Timed execution is reported only for workloads that pass this validation. This separates correctness from performance, ensuring that execution-time and memory comparisons are made only across semantically equivalent workloads.

\subsubsection{Timing and Memory Measurement}
\label{subsubsec:timing_memory}

All compared backends are accessed through a unified interface with common stages for preparation, parity checking, inference, and training. The shared scalar objective is the mean of single-qubit Pauli-$Z$ expectations under deterministic parameter bindings. Inference measures the evaluation of this objective. Training measures one gradient-evaluation step on the same objective, without an optimizer update.

All timed RQ2 runs are GPU-only. Only workloads that pass parity validation are timed. Each measurement uses a fixed warmup-and-repeat protocol with device synchronization around the timed region, and we report mean and standard deviation across repeats. In the strict RQ2 sweep, training uses a matched central finite-difference policy (FD-matched) for all trainable backends. Separate auxiliary runs use native training paths where available: VQCSim uses PyTorch autograd, PennyLane uses a gradient-capable QNode path when supported, and TensorCircuit uses native autodiff with finite-difference fallback when necessary. Qiskit training is finite-difference, and CUDA-Q is evaluated only in inference mode because the evaluated RQ2 adapter lacks a compatible training path~\cite{bergholm2018pennylane,qiskit2024,tensorcircuit2023,cudaquantum}. For fairness, PennyLane QNodes are instantiated once during preparation and reused across timing repeats.

Peak per-process VRAM and GPU utilization statistics are sampled concurrently through NVML when available, and monitor availability is recorded explicitly. To keep the sweep robust, the protocol also includes per-job timeouts and GPU memory guards before and after preparation. Failures are explicitly recorded as unsupported-mode, semantic, resource, or runtime-internal outcomes rather than silently discarded. Each run stores a reproducibility manifest containing software versions, preprocessing and transpilation settings, parity tolerances, and execution-environment information~\cite{quetschlich2023mqtbench,qiskit2024}.

\subsubsection{Ablation and Mechanism Analysis}
\label{subsubsec:ablation_setup}
RQ3 isolates the sources of VQCSim's advantage through two complementary analyses on a training-only workload slice drawn from the same variational families as RQ2.

\textit{Cross-framework fairness test.} To determine how much of VQCSim's training speedup is explained by its native autograd path rather than by the compile-once execution path itself, we re-run all framework--circuit--batch combinations under a matched central finite-difference gradient policy. This forces every backend to use the same differentiation mechanism, so that residual speedup differences can be attributed solely to the execution path.

\textit{Internal ablation.} We construct three degraded variants of VQCSim, each disabling exactly one mechanism: (1)~\textit{no autograd} replaces native PyTorch backpropagation with the same central finite-difference policy used in the fairness test; (2)~\textit{no compile-once} re-parses and re-lowers the OpenQASM circuit on every forward call instead of reusing the cached gate sequence; (3)~\textit{no batch vectorization} loops over individual batch elements sequentially rather than applying gates to the full $B \times 2^n$ state tensor. Each variant is timed against the full VQCSim path at batch sizes $1$ and $32$ using the same warmup-and-repeat protocol as RQ2. Significance is assessed with paired Wilcoxon signed-rank tests.

\subsubsection{Backend Selector Evaluation}
\label{subsubsec:selector_setup}
RQ4 evaluates whether the empirical findings from RQ2 and RQ3 can be converted into an automatic backend-selection rule. We train a lightweight selector on the runtime and memory records collected in RQ2, using observable pre-run features such as qubit count, batch size, execution mode, circuit depth, gate count, and parameter count. The selector is evaluated on a held-out family-size split disjoint from the training data. We report top-1 oracle agreement, meaning how often it selects the oracle-best backend, regret, meaning the runtime cost relative to the oracle choice, and false-feasible rate, meaning how often it incorrectly predicts VQCSim to fit and run.

\section{Experimental Results}

This section first establishes applicability and correctness, then quantifies performance and its memory tradeoff, next analyzes the mechanisms behind the observed behavior through scaling and ablation evidence, and finally converts these findings into a practical backend-selection rule.

\subsection{Applicability and Correctness}
\label{subsec:rq1_results}

VQCSim is broadly applicable across the tested static MQT Bench population, but its practical boundary is validated semantic agreement rather than path construction.

We evaluate applicability over algorithm-level MQT Bench circuits for all discovered valid static sizes up to 20~\cite{quetschlich2023mqtbench}, following the preprocessing and validation protocol described in \autoref{subsec:experimental_setup}. The study covers 463 static circuits from 31 benchmark families. All 463 circuits remain in scope after preprocessing, and lowering succeeds for all 463/463 circuits, showing that a compile-once Torch-native path can be constructed for every tested static circuit. External parity validation succeeds for 406 circuits, yielding an end-to-end validation rate of \(87.7\%\). Among the remaining 57 cases, 37 are semantic mismatches, and 20 are out-of-memory (OOM) runs in the reference checker. Restricting attention to completed semantic checks leaves 443 circuits, of which 406 agree with the reference, for a semantic agreement rate of \(91.6\%\).

Support is broad but not uniform. Twenty-two of the 31 benchmark families achieve \(100\%\) validated support over the tested range, and all parametrized circuits are supported. The unsupported cases are highly localized: \texttt{grover} and \texttt{qwalk} account for 30 of the 37 semantic failures, or about \(81\%\), with first failures at size 6. The remaining mismatches are limited to \texttt{ae} (4), \texttt{multiplier} (1), \texttt{rg\_qft\_} \texttt{multiplier} (1), and \texttt{shor} (1). This concentration is consistent with the structural diagnostics. Supported circuits have a median gate count of 102.5 and a median two-qubit gate count of 35.5, whereas semantic-failure circuits have medians of 57{,}790 and 20{,}886, respectively, indicating that the current boundary is defined mainly by a small set of deep, structurally heavy families rather than by broad incompatibility across MQT Bench.

The 20 out-of-memory cases are validation-limited rather than lowering-limited, since lowering succeeds, but the strict Qiskit parity oracle cannot complete within the available host-memory budget for some larger \texttt{qft}, \texttt{qftentangled}, \texttt{hhl}, and \texttt{ae} circuits.

\subsection{Performance and Memory Tradeoff}
\label{subsec:rq2_results}

VQCSim is a throughput-oriented execution path whose advantage grows with workload aggregation and remains stable across the tested multi-GPU set, but this speedup is purchased at the cost of substantially higher GPU memory use.

\begin{figure}[htbp]
    \centering
    \includegraphics[width=0.85\linewidth]{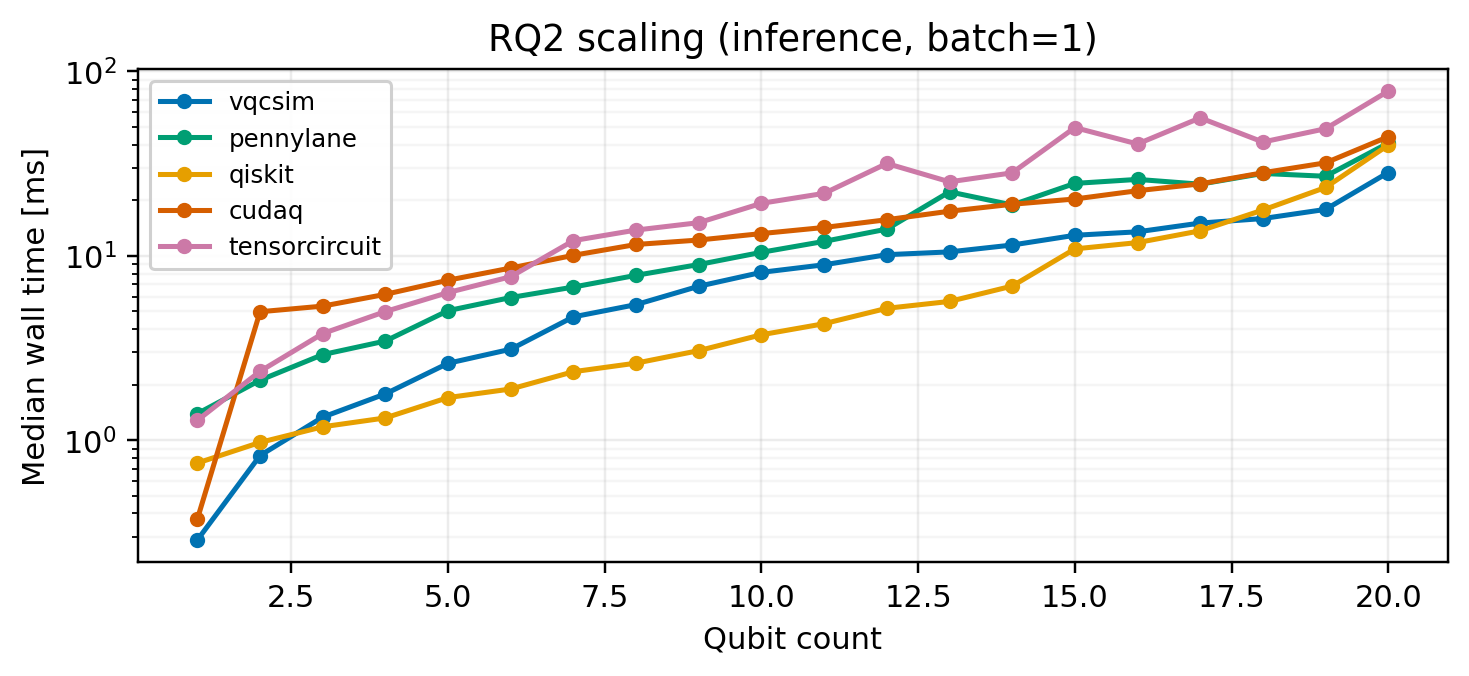}
    \includegraphics[width=0.85\linewidth]{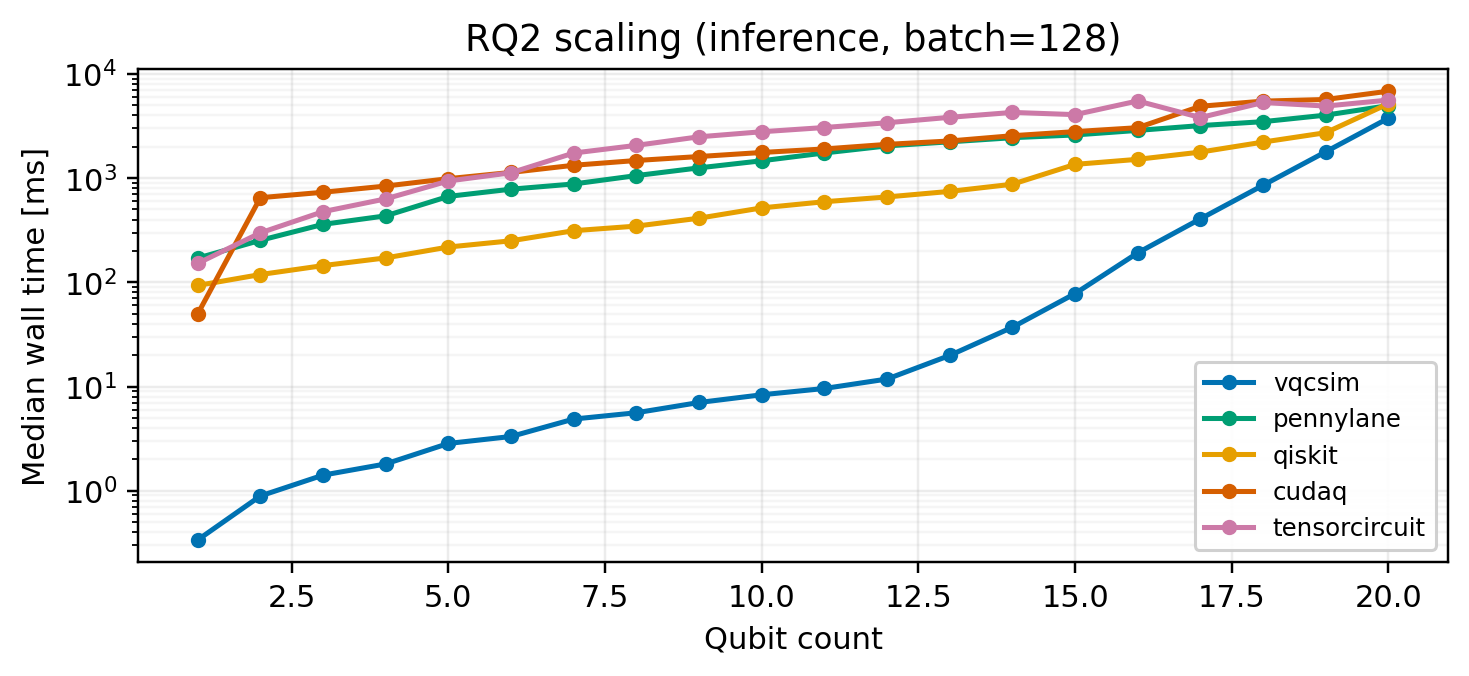}
    \caption{Median inference wall time versus qubit count for representative small and large batch regimes. The top plot shows batch size \(1\), where VQCSim is competitive, but the regime remains mixed. The bottom plot shows batch size \(128\), where VQCSim operates in a clearly lower wall-time regime than the generic baselines across most of the tested qubit range. The wall-time axis is logarithmic.}
    \label{fig:rq2_scaling_inference}
\end{figure}

\autoref{fig:rq2_scaling_inference} first shows the basic crossover on the main RTX~5080 evaluation platform. At batch size \(1\), VQCSim is competitive, but the regime remains mixed. At batch size \(128\), it moves into a clearly lower wall-time regime than the generic baselines over most of the tested qubit range. \autoref{tab:rq2_crossgpu_speedups} shows that this behavior generalizes across the tested GPUs rather than arising from a single-device effect. Its main outcome is that VQCSim's advantage is primarily a batch-regime effect: low-batch inference remains mixed, but from batch \(8\) onward the results increasingly favor VQCSim. Training is even more favorable to specialization, with native training preferring VQCSim throughout the tested range, while matched-gradient training again shows a mixed low-batch regime followed by clear VQCSim advantages at moderate and large batch sizes.

\begin{table}[htbp]
    \centering
    \caption{Equal-weighted cross-GPU median speedup of each shared baseline relative to VQCSim on the strict common subset, reported as baseline runtime divided by VQCSim runtime. Values below \(1\) mean the baseline is faster than VQCSim, while values above \(1\) favor VQCSim. Red marks losses, yellow marks near-parity or small wins, and darker green marks larger VQCSim advantages.}
    \label{tab:rq2_crossgpu_speedups}
    \vspace*{0.5em}
    \setlength{\tabcolsep}{3pt}
    \renewcommand{\arraystretch}{1.08}
    \resizebox{\linewidth}{!}{%
    \begin{tabular}{llrrrr}
        \toprule
        Regime & Framework & Batch \(1\) & Batch \(8\) & Batch \(32\) & Batch \(128\) \\
        \midrule
        Native inference         & PennyLane
            & \cellcolor{lossred}0.52
            & \cellcolor{winmid}3.97
            & \cellcolor{windark!75}\textcolor{white}{14.85}
            & \cellcolor{windark!90}\textcolor{white}{57.07} \\
        Native inference         & Qiskit
            & \cellcolor{lossred}0.39
            & \cellcolor{winmid}3.48
            & \cellcolor{windark!70}\textcolor{white}{11.83}
            & \cellcolor{windark!85}\textcolor{white}{41.97} \\
        Native training          & PennyLane
            & \cellcolor{winlight}2.17
            & \cellcolor{windark!75}\textcolor{white}{15.89}
            & \cellcolor{windark!90}\textcolor{white}{53.52}
            & \cellcolor{windark!95}\textcolor{white}{244.22} \\
        Native training          & Qiskit
            & \cellcolor{winmid}6.51
            & \cellcolor{windark!80}\textcolor{white}{38.88}
            & \cellcolor{windark!92}\textcolor{white}{84.32}
            & \cellcolor{windark!95}\textcolor{white}{310.55} \\
        Fair FD-matched training & PennyLane
            & \cellcolor{lossred}0.59
            & \cellcolor{winmid}4.20
            & \cellcolor{windark!78}\textcolor{white}{17.68}
            & \cellcolor{windark!92}\textcolor{white}{82.97} \\
        Fair FD-matched training & Qiskit
            & \cellcolor{lossred}0.41
            & \cellcolor{winmid}3.09
            & \cellcolor{windark!72}\textcolor{white}{12.60}
            & \cellcolor{windark!90}\textcolor{white}{71.85} \\
        \bottomrule
    \end{tabular}}
\end{table}

\begin{figure}[htbp]
    \centering
    \includegraphics[width=\linewidth]{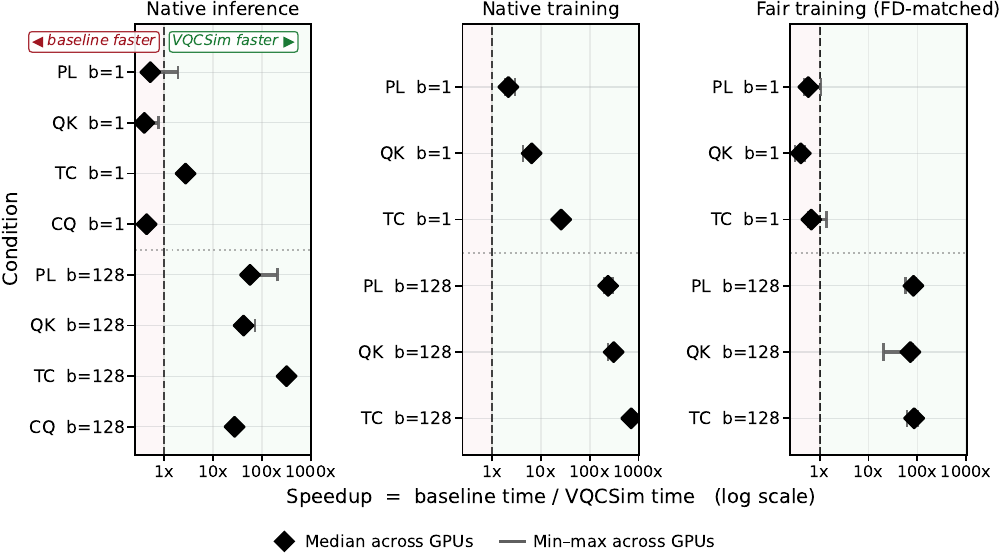}
    \caption{Cross-GPU generalization of VQCSim speedups. Rows are baseline--batch conditions at \(b=1\) and \(b=128\); panels report native inference, native training, and FD-matched training (matched finite-difference gradients). Diamonds show the median speedup across the five tested GPUs and whiskers the min--max range; tight ranges indicate the regime pattern is device-independent. CUDA-Q (CQ) is inference-only (Section~4.2); per-GPU values are included in the released artifacts.}
    \label{fig:cross-gpu-generalisation}
\end{figure}

\autoref{fig:cross-gpu-generalisation} then shows that this crossover is not tied to one device. Across the tested GPU platforms, the same regime-level pattern appears consistently: native inference is mixed for the smallest jobs, but for larger batches, VQCSim moves into a clearly better throughput regime; native training favors VQCSim throughout; and fairness-controlled matched training is again mixed at batch \(1\) but clearly favorable at moderate and large batch sizes. \autoref{tab:rq2_native_headlines} summarizes the aggregate effect: the pooled median speedup is \(4.49\times\) for native inference, \(26.78\times\) for native training, and \(3.31\times\) for fair FD-matched training. Thus, the large native training gain is partly amplified by the full differentiation stack, but a clear simulator-level advantage remains under matched finite differences.

\begin{table}[htbp]
    \centering
    \caption{Headline pooled speedups on the strict common cross-GPU subset. Values are baseline runtime divided by VQCSim runtime, so values greater than \(1\) favor VQCSim.}
    \label{tab:rq2_native_headlines}
    \vspace*{0.5em}
    \resizebox{\linewidth}{!}{%
    \begin{tabular}{llc}
        \toprule
        Workload & Aggregate view & Median speedup \\
        \midrule
        Native inference         & pooled common-subset median & \(4.49\times\) \\
        Native training          & pooled common-subset median & \(26.78\times\) \\
        Fair FD-matched training & pooled common-subset median & \(3.31\times\) \\
        \bottomrule
    \end{tabular}}
\end{table}

\begin{figure}[t]
    \centering
    \includegraphics[width=\linewidth]{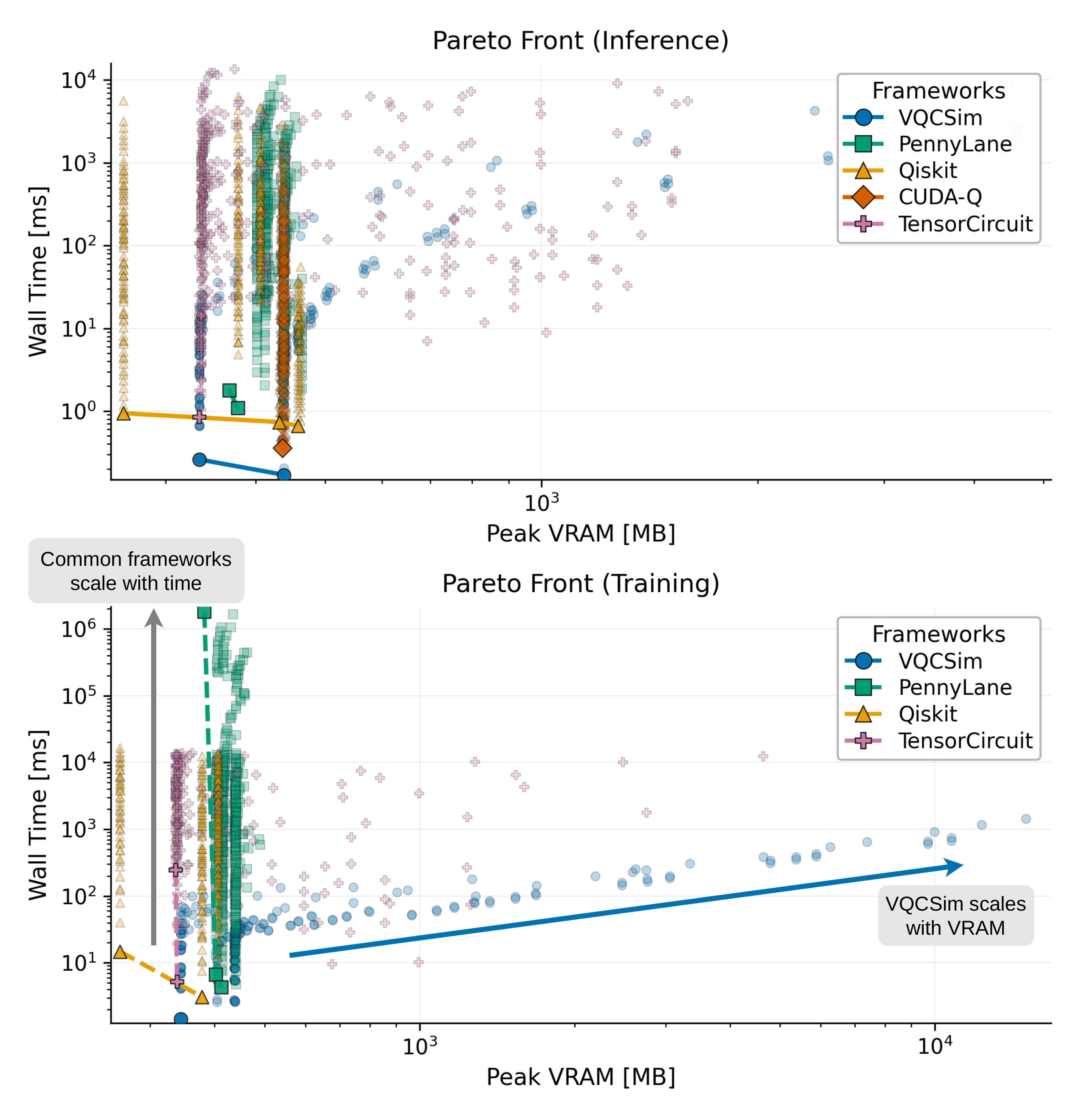}
    \caption{Pareto view of execution time versus peak GPU VRAM for VQCSim and generic simulators. Lower-left is better. The frontiers summarize the feasible speed--memory tradeoff in inference and training. VQCSim increasingly occupies the low-latency region as workload size grows, but often at a higher VRAM cost, especially in training.}
    \label{fig:rq2_pareto}
\end{figure}

The tradeoff for this throughput advantage is memory, and \autoref{fig:rq2_pareto} makes that tradeoff explicit. In both inference and training, VQCSim increasingly occupies the low-latency region, but it does so by moving much farther along the VRAM axis than generic frameworks do. This is the central performance interpretation of RQ2: VQCSim does not win by reducing both time and memory simultaneously. It wins by trading memory for throughput. Over successful runs in the common multi-GPU subset, native inference peak VRAM is about \(4.39\text{--}4.57\)~GB for VQCSim, versus \(0.31\text{--}0.46\)~GB for PennyLane or Qiskit. In native training, VQCSim rises to \(15.0\text{--}48.1\)~GB, whereas PennyLane and Qiskit remain around \(0.26\text{--}0.49\)~GB. Under fair matched training, VQCSim's successful-run VRAM drops to \(0.51\text{--}2.40\)~GB, but remains above the generic baselines.

The aggregate speedups above could still hide family-specific variation. To make the performance result more interpretable at the benchmark level, \autoref{tab:headline_family_winrate} reports family-wise VQCSim win rates on the main RTX~5080 platform across all sizes and batch settings. The pattern is broad rather than localized: VQCSim wins nearly all training instances against the generic baselines, while inference remains more mixed, especially against Qiskit and CUDA-Q on \texttt{qaoa}. The main conclusion of RQ2 is therefore not driven by a single family but reflects a consistent advantage across the tested set of variational workloads.

Taken together, these results show that VQCSim is least compelling for tiny, low-batch workloads but becomes increasingly advantageous as workload aggregation increases, provided sufficient memory headroom is available. These conditions are typical of many hybrid-QML training and design-space-exploration workflows in the low-qubit regime, where repeated batched circuit evaluations dominate end-to-end cost.

\definecolor{winlow}{RGB}{232,245,233}
\definecolor{winmid}{RGB}{165,214,167}
\definecolor{winhigh}{RGB}{76,175,80}
\definecolor{wintop}{RGB}{46,125,50}

\begin{table}[t]
    \centering
    \caption{Family-wise VQCSim win rates (\%) on RTX~5080 across sizes \(1\ldots20\) and batch sizes \(\{1,8,32,128\}\). Columns compare VQCSim against Qiskit (QK), PennyLane (PL), TensorCircuit (TC), and CUDA-Q (CQ). Rows are split into inference (Inf.) and training (Tr.). Darker green indicates a higher win rate. CUDA-Q training is unavailable in the experiment setting (see \autoref{subsec:experimental_setup}).}
    \label{tab:headline_family_winrate}
    \vspace*{0.5em}
    \setlength{\tabcolsep}{3pt}
    \renewcommand{\arraystretch}{1.08}
    \resizebox{0.55\columnwidth}{!}{%
    \begin{tabular}{llcccc}
        \toprule
        Family & Mode & QK & PL & TC & CQ \\
        \midrule
        \multirow{2}{*}{\texttt{qaoa}}
            & Inf. & \cellcolor{winlow}73.8 & \cellcolor{wintop!85}\textcolor{white}{100} & \cellcolor{wintop!85}\textcolor{white}{100} & \cellcolor{winlow}65.0 \\
            & Tr.  & \cellcolor{winmid}87.8 & \cellcolor{wintop!85}\textcolor{white}{100} & \cellcolor{wintop!85}\textcolor{white}{100} & -- \\
        \midrule
        \multirow{2}{*}{\texttt{qnn}}
            & Inf. & \cellcolor{winmid}82.5 & \cellcolor{wintop!85}\textcolor{white}{100} & \cellcolor{wintop!85}\textcolor{white}{100} & \cellcolor{winlow}75.0 \\
            & Tr.  & \cellcolor{wintop!85}\textcolor{white}{100} & \cellcolor{wintop!85}\textcolor{white}{100} & \cellcolor{wintop!85}\textcolor{white}{100} & -- \\
        \midrule
        \multirow{2}{*}{\texttt{vqe\_real\_amp}}
            & Inf. & \cellcolor{winmid}85.0 & \cellcolor{wintop!85}\textcolor{white}{100} & \cellcolor{wintop!85}\textcolor{white}{100} & \cellcolor{winmid}77.5 \\
            & Tr.  & \cellcolor{wintop!85}\textcolor{white}{100} & \cellcolor{wintop!85}\textcolor{white}{100} & \cellcolor{wintop!85}\textcolor{white}{100} & -- \\
        \midrule
        \multirow{2}{*}{\texttt{vqe\_su2}}
            & Inf. & \cellcolor{winmid}82.5 & \cellcolor{wintop!85}\textcolor{white}{100} & \cellcolor{wintop!85}\textcolor{white}{100} & \cellcolor{wintop!85}\textcolor{white}{100} \\
            & Tr.  & \cellcolor{wintop!85}\textcolor{white}{100} & \cellcolor{wintop!85}\textcolor{white}{100} & \cellcolor{wintop!85}\textcolor{white}{100} & -- \\
        \midrule
        \multirow{2}{*}{\texttt{vqe\_two\_local}}
            & Inf. & \cellcolor{winmid}85.5 & \cellcolor{winhigh}93.4 & \cellcolor{wintop!85}\textcolor{white}{100} & \cellcolor{wintop!85}\textcolor{white}{100} \\
            & Tr.  & \cellcolor{wintop!85}\textcolor{white}{100} & \cellcolor{wintop!85}\textcolor{white}{100} & \cellcolor{wintop!85}\textcolor{white}{100} & -- \\
        \midrule
        \multirow{2}{*}{\textbf{Overall}}
            & Inf. & \cellcolor{winmid}81.8 & \cellcolor{winhigh}98.7 & \cellcolor{wintop!85}\textcolor{white}{100} & \cellcolor{winmid}85.3 \\
            & Tr.  & \cellcolor{winhigh}97.6 & \cellcolor{wintop!85}\textcolor{white}{100} & \cellcolor{wintop!85}\textcolor{white}{100} & -- \\
        \bottomrule
    \end{tabular}}
\end{table}

\subsection{Mechanism Analysis and Ablation}
\label{subsec:rq3_results}

We now turn from \textit{what} performance regime VQCSim occupies to \textit{why} that regime arises. The evidence in \autoref{fig:rq2_pareto} already suggests that VQCSim and the generic frameworks do not hit the same practical wall: VQCSim increasingly trades VRAM for lower latency, whereas the generic stacks drift upward in wall time before memory becomes dominant. RQ3, therefore, asks two related questions: which workload and circuit components dominate runtime and memory in each backend class, and which internal mechanisms create VQCSim's throughput advantage.

At the backend level, the dominant mechanisms differ clearly between VQCSim and the generic frameworks. In VQCSim, runtime is primarily determined by circuit width and structural complexity. Qubit count is the clearest first-order driver because it sets the underlying state size, but it is not sufficient on its own: depth, total gate count, two-qubit gate density, parameter count, and gate-type mix all shift the execution constant factors. In particular, circuits with higher two-qubit density or a larger share of more expensive gate applications incur noticeably higher runtime at the same qubit count. VQCSim memory usage is likewise dominated by more than the raw state alone. The base state representation grows exponentially with qubit count, while parameter count, gate volume, and depth increase the amount of backward-path state that must be retained. The first practical limit in VQCSim is therefore usually memory pressure during large training workloads.

The generic frameworks exhibit different scaling profiles, dominated by batch growth, framework orchestration, and gradient-stack overhead, especially during training. Circuit descriptors still matter, but their effect is often mediated by repeated framework-level dispatch and objective evaluations rather than solely by the underlying state evolution, making these stacks much more batch-amplifying than VQCSim. Their GPU memory usage on successful runs remains comparatively low, and many failures are not accompanied by substantial VRAM growth, so the practical limit is usually runtime rather than memory. Qiskit and PennyLane are therefore primarily runtime-limited, while TensorCircuit remains the main mixed case, showing both early timeout behavior and occasional high-end memory spikes.

This mechanism summary explains the different failure modes already visible in \autoref{fig:rq2_pareto}. In VQCSim, width determines the simulation regime, and structural complexity shifts the frontier, so circuits with the same qubit count can still achieve noticeably different runtime and VRAM usage. The resulting high-end limit is therefore primarily a VRAM ceiling during native training. In generic frameworks, the first barrier is often overhead and runtime growth, so timeouts occur before memory exhaustion on many workloads. Thus, the specialized and generic stacks do not merely differ in degree of scaling, but in which resource becomes active first.

We next isolate how much of VQCSim's training advantage stems from its execution path and how much from its native differentiation stack. \autoref{tab:ablation_fairness_pooled} provides the cross-framework fairness control: when all frameworks are forced to use the same finite-difference gradient policy, VQCSim still achieves a pooled median training speedup of \(2.15\times\) on \(n=446\). This result is much smaller than the native-training headline from \autoref{tab:rq2_native_headlines}, but it remains significant and therefore shows that VQCSim retains a genuine simulator-level advantage even after removing gradient-policy differences.

\begin{table}[t]
    \centering
    \caption{Fairness-controlled training speedup with matched finite differences across frameworks. Ratios are baseline runtime divided by VQCSim runtime, so values greater than \(1\) favor VQCSim. All significance tests are paired Wilcoxon signed-rank tests against the neutral value \(1.0\).}
    \label{tab:ablation_fairness_pooled}
    \vspace*{0.5em}
    \resizebox{0.8\linewidth}{!}{%
    \begin{tabular}{lrrrr}
        \toprule
        Scope & \(n\) & Median speedup & \(p\) & \(r_{\mathrm{bc}}\) \\
        \midrule
        Overall & 446 & \(2.148\pm0.349\)x & \(1.23\times10^{-55}\) & 0.859 \\
        PennyLane & 185 & \(2.239\pm0.417\)x & \(4.07\times10^{-28}\) & 0.932 \\
        Qiskit & 140 & \(1.036\pm0.150\)x & \(2.14\times10^{-3}\) & 0.299 \\
        TensorCircuit & 121 & \(4.206\pm0.356\)x & \(1.35\times10^{-21}\) & 1.000 \\
        \bottomrule
    \end{tabular}}
    \vspace*{-1.5em}
\end{table}

Having controlled for gradient policy externally, we then ablate VQCSim internally to determine which of its own design choices actually create the speedup. \autoref{tab:ablation_internal} shows a clear ordering. Removing native autograd produces by far the largest slowdown, with a pooled median of \(27.63\times\). Removing compile-once lowering causes a much smaller but highly consistent slowdown of \(1.91\times\). Removing batch vectorization is strongly regime-dependent: it is not significant at batch \(1\), but becomes catastrophic at batch \(32\), where the slowdown reaches \(25.95\times\). The mechanism picture is therefore consistent across the internal and external analyses. Native autograd is the dominant contributor to VQCSim's training advantage; compile-once lowering provides a stable secondary gain; and vectorized execution becomes critical once workloads are sufficiently batched.

\begin{table}[t]
    \centering
    \caption{Internal VQCSim ablation. Ratios are the ablated runtime divided by the full VQCSim runtime; values greater than \(1\) indicate the ablated version is slower.}
    \label{tab:ablation_internal}
    \vspace*{0.5em}
    \resizebox{\linewidth}{!}{%
    \begin{tabular}{llrrrr}
        \toprule
        Ablation & Batch & \(n\) & Median slowdown & \(p\) & \(r_{\mathrm{bc}}\) \\
        \midrule
        No autograd & \(1\) & 57 & \(32.294\pm20.789\)x & \(5.14\times10^{-11}\) & 1.000 \\
        No autograd & \(32\) & 50 & \(25.512\pm20.458\)x & \(1.78\times10^{-15}\) & 1.000 \\
        No autograd & pooled & 107 & \(27.628\pm16.863\)x & \(2.73\times10^{-19}\) & 1.000 \\
        \midrule
        No compile-once & \(1\) & 60 & \(1.936\pm0.036\)x & \(1.71\times10^{-11}\) & 0.999 \\
        No compile-once & \(32\) & 55 & \(1.905\pm0.166\)x & \(2.03\times10^{-10}\) & 0.986 \\
        No compile-once & pooled & 115 & \(1.914\pm0.048\)x & \(1.95\times10^{-20}\) & 0.996 \\
        \midrule
        No batch vectorization & \(1\) & 60 & \(0.825\pm0.023\)x & \(9.12\times10^{-1}\) & 0.016 \\
        No batch vectorization & \(32\) & 55 & \(25.947\pm0.466\)x & \(1.11\times10^{-10}\) & 1.000 \\
        No batch vectorization & pooled & 115 & \(1.861\pm5.808\)x & \(1.08\times10^{-11}\) & 0.730 \\
        \bottomrule
    \end{tabular}}
    \vspace*{-1.5em}
\end{table}

In summary, VQCSim is fast not because it escapes the underlying statevector scaling regime, but because it removes the framework and differentiation overhead that dominates generic execution at moderate and large batch sizes.  This also explains why the relative speedup shrinks with qubit count: each additional qubit doubles the number of states that every simulator must process, so the eliminated overhead becomes a smaller fraction of the total runtime. The cost of this strategy is that the high-end limit shifts from runtime to memory; exact thresholds remain hardware- and timeout-dependent, so the claim holds at the mechanism level rather than the threshold level.

\subsection{Practical Backend-Selection Rule}
\label{sec:regime_map}
The results of RQ1--RQ3 support a simple constrained backend-selection rule: filter first by support, then by memory feasibility, and only then by expected runtime.

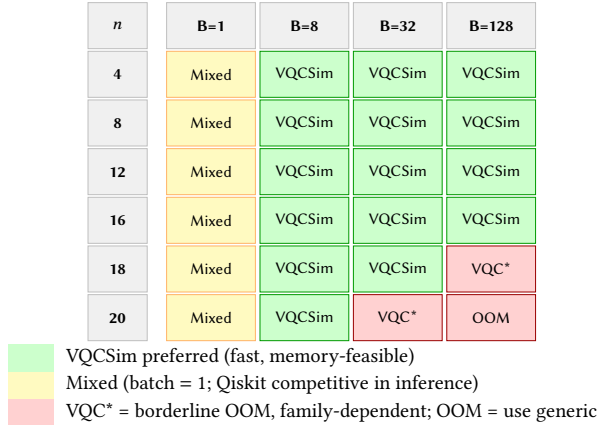
\begin{figure}[t]
\centering
\resizebox{0.7\linewidth}{!}{%
\begin{tikzpicture}[
  font=\sffamily\footnotesize,
  cell/.style={rectangle, minimum width=1.35cm, minimum height=0.7cm, align=center},
  vqc/.style={cell, fill=green!20, draw=green!60!black},
  mix/.style={cell, fill=yellow!30, draw=orange!60},
  oom/.style={cell, fill=red!18, draw=red!60!black},
  hdr/.style={cell, fill=gray!12, draw=gray!50, font=\sffamily\footnotesize\bfseries},
]
\def\dx{1.42}
\def\dy{0.74}
\foreach \col/\lbl in {1/{B=1},2/{B=8},3/{B=32},4/{B=128}}
  \node[hdr] at (\col*\dx, 0) {\lbl};
\node[hdr,minimum width=0.9cm] at (0,0) {$n$};
\foreach \row/\lbl in {1/4,2/8,3/12,4/16,5/18,6/20}
  \node[hdr,minimum width=0.9cm] at (0,-\row*\dy) {\lbl};
\foreach \row in {1,2,3,4}
  \foreach \col in {2,3,4}
    \node[vqc] at (\col*\dx,-\row*\dy) {VQCSim};
\foreach \row in {1,2,3,4,5,6}
  \node[mix] at (1*\dx,-\row*\dy) {Mixed};
\foreach \col in {2,3}
  \node[vqc] at (\col*\dx,-5*\dy) {VQCSim};
\node[oom] at (4*\dx,-5*\dy) {VQC*};
\node[vqc] at (2*\dx,-6*\dy) {VQCSim};
\node[oom] at (3*\dx,-6*\dy) {VQC*};
\node[oom] at (4*\dx,-6*\dy) {OOM};
\end{tikzpicture}}

\resizebox{0.9\linewidth}{!}{%
\small
\begin{tabular}{@{}ll@{}}
\cellcolor{green!20}\hspace{8pt} & VQCSim preferred (fast, memory-feasible)\\[1pt]
\cellcolor{yellow!30}\hspace{8pt} & Mixed (batch \(=1\); Qiskit competitive in inference)\\[1pt]
\cellcolor{red!18}\hspace{8pt} & VQC* = borderline OOM, family-dependent; OOM = use generic\\
\end{tabular}}
\vspace*{1em}
\caption{Hardware-aware regime map (RTX~5080, 16\,GB). Rows correspond to qubit count \(n\); columns correspond to batch size \(B\). VQCSim is preferred in the green region. The batch \(=1\) column is mixed because Qiskit remains competitive for tiny inference jobs. The OOM boundary moves by about one to two qubits across benchmark families due to circuit-structure effects.}
\label{fig:hardware-map}
\end{figure}

The decision procedure has three required stages, one built-in policy guard, and one optional refinement step. First, \textit{support filtering} removes any backend that cannot legally execute the requested job; for VQCSim, this boundary is defined directly by the static-circuit applicability results in \autoref{subsec:rq1_results}. Second, \textit{memory-feasibility filtering} removes backends that are unlikely to fit within the available memory budget. Because the mechanism analysis in \autoref{subsec:rq3_results} shows a qubit-only rule is insufficient for VQCSim, a conservative estimate combines a base state-memory term with an overhead term predicted from structural descriptors (depth, gate volume, parameter count, two-qubit density), plus an additional guard in stress training regimes (\(b\ge 32\), large \(q\)). Third, \textit{performance selection} minimizes expected runtime over the feasible subset. A built-in low-batch inference guard prevents over-selection of VQCSim in \(b=1\) crossover regimes by preferring a feasible generic alternative, and when top candidates are near-tied, an optional short online probe can replace model-only ranking. The released \texttt{vqcsim-oracle} implements this procedure and returns a machine-readable report with the backend decision, alternatives, rejection rationale, fallback actions, and provenance. \autoref{fig:hardware-map} summarizes the resulting rule on the main RTX~5080 platform: the VQCSim-preferred region is broad at moderate and large batch sizes, mixed at batch \(1\), and narrows near the high-end memory boundary.

When evaluated on previously unseen family-size cases, the selector performs well. Its choices are usually oracle-consistent, its typical regret is minimal, and false-feasible recommendations remain rare. The residual errors are concentrated in exactly the regimes identified earlier in \autoref{subsec:rq2_results} and \autoref{subsec:rq3_results}: low-batch inference, where VQCSim's throughput advantage is weakest, and a small number of large training cases near the VQCSim memory boundary. In other words, the selector is strongest in the same throughput-oriented regimes where VQCSim itself is strongest, and weakest near the known crossover and feasibility boundaries.

We also evaluated cross-hardware transfer by training the selector on RTX~5080 records and applying the same fixed model to four additional GPU targets. \autoref{tab:oracle-transfer-5080} shows that the learned rule transfers well across devices: top-1 accuracy remains high, regret stays near-optimal, and false-feasible rates remain low, with most errors concentrated in known hard regimes such as low-batch inference and boundary cases. This indicates that the selector captures portable regime distinctions rather than overfitting to a single GPU. In practice, VQCSim is preferred when the circuit is supported and memory-feasible, especially for moderate- and large-batch workloads, whereas generic backends remain preferable for tiny inference jobs and memory-constrained cases.

\section{Conclusions}
\label{sec:conclusion}

We presented VQCSim, a compile-once PyTorch-native execution path for supported static quantum circuits under ideal statevector simulation. Across 463 static MQT Bench circuits, the lowering stage succeeds in all cases, and end-to-end validated support reaches \(87.7\%\); among circuits for which the reference check completes, semantic agreement reaches \(91.6\%\). In the common multi-GPU comparison on variational circuit families, VQCSim delivers pooled median speedups of \(4.49\times\) for native inference and \(26.78\times\) for native training, while retaining a \(3.31\times\) pooled median advantage under fair FD-matched training. The largest gains appear in moderate- and large-batch regimes, whereas tiny low-batch inference workloads remain more mixed.

Component-level analysis shows that the two execution paths follow different descriptor-driven scaling regimes. In VQCSim, cost is mainly determined by qubit count and circuit structure, with native PyTorch autograd providing the largest speedup, and compile-once caching and batch vectorization adding further gains. In generic frameworks, batch size, framework-level orchestration, and gradient overhead dominate much more strongly, especially during training. For supported circuits, this makes the VQC component much more practical inside hybrid pipelines by trading higher GPU VRAM demand for much lower wall time while preserving circuit behavior. At the high end, VQCSim becomes memory-limited, whereas generic frameworks are often limited by runtime overhead.

\begin{table}[t]
\centering
\caption{Cross-hardware transfer of the RTX~5080-trained VQCSim oracle to other GPUs. Cases differ slightly across target GPUs because the evaluation includes only target-side benchmark points with completely comparable recorded outcomes.}
\label{tab:oracle-transfer-5080}
\vspace*{1em}
\resizebox{0.9\linewidth}{!}{%
\begin{tabular}{lrrrr}
\toprule
Target GPU & Cases & Top-1 Acc. (\%) & Regret$_{95}$ & False-feasible (\%) \\
\midrule
NVIDIA RTX 4000 Ada & 636 & 96.70 & 1.000 & 1.41 \\
NVIDIA RTX 4500 Ada & 637 & 97.65 & 1.000 & 0.94 \\
NVIDIA RTX A5000 & 637 & 96.39 & 1.000 & 0.94 \\
NVIDIA RTX A6000 & 638 & 91.07 & 1.191 & 0.78 \\
\bottomrule
\end{tabular}}
\vspace{-1.5em}
\end{table}

For practitioners building hybrid QML pipelines, this tradeoff is practically important. As an illustration, the pooled median native-training speedup of \(26.78\times\) means that, under comparable conditions, a workload taking about \(26.8\) hours on a generic framework would finish in about \(1\) hour with VQCSim. For supported static circuits at the low-to-moderate qubit counts common in current QML, this can turn quantum layers from a prohibitive iteration bottleneck into a usable component of repeated training, ablation, and model-selection loops, provided memory remains feasible.

In summary, the results turn backend choice for static QML circuits from trial and error into a practical rule, which we encode in \texttt{vqcsim-oracle} as a CAD aid for hybrid QML workflows. The present results are scoped to ideal statevector simulation of small static circuits on the evaluated GPU platforms; future work includes broader cross-hardware calibration, explicit VRAM-budget-aware feasibility checks, extension beyond the noise-free setting, and follow-up on the small set of structurally heavy families that account for most remaining mismatches.

\begin{acks}
This work was supported by the Brno University of Technology internal project FIT-S-26-9011 and the Czech Science Foundation grant 26-22525M EvoML-EDA. Computational resources were provided by the e-INFRA CZ project (ID:90254), supported by the Ministry of Education, Youth and Sports of the Czech Republic.
This work was also supported in part by the NYUAD Center for Quantum and Topological Systems (CQTS), funded by Tamkeen under the NYUAD Research Institute grant CG008, and the Center for Cyber Security (CCS), funded by Tamkeen under the NYUAD Research Institute Award G1104.
The authors used ChatGPT and Claude for copy-editing, reviewed all AI-assisted text, and take full responsibility for the content.
\end{acks}

\bibliographystyle{ACM-Reference-Format}
\bibliography{sample-base}

\end{document}